\documentclass[11pt,letterpaper]{article}
\pdfoutput=1
\usepackage{ifthen}
\usepackage[utf8]{inputenc}
\usepackage{mathtools}
\usepackage{amscd}
\usepackage{amsmath}
\usepackage{slashed}
\usepackage{bm}
\usepackage{jheppub} 
\usepackage{caption} 
\usepackage{subcaption}
\newboolean{proc}
\setboolean{proc}{false}

\title{Fits to measurements of rare heavy flavour decays}

\author[a]{Ben Allanach}
\affiliation[a]{DAMTP, University of Cambridge, Wilberforce Road, Cambridge, 
  CB3 0WA, United Kingdom}
\emailAdd{B.C.Allanach@damtp.cam.ac.uk}
\abstract{This write-up is intended to form part of the proceedings for
  Lepton-Photon 2023.
  We review the decays $b\rightarrow c \ell \bar \nu_\ell$
  (where $\ell \in \{e,\mu,\tau\}$)
  as well as $b\rightarrow s \mu^+ \mu^-$ and $b\rightarrow s e^+ e^-$, giving
  the  
  current state-of-the-art in terms of measurements. We review fits to such
  data of new physics 
  weak effective field theory operators, before closing with 
  interpretations in terms of simplified TeV-scale field theories.}
  

\begin{document}
\maketitle
\flushbottom
\ifthenelse{\boolean{proc}}{}
{\section{$b\rightarrow c \tau \nu$ decays\label{sec:intro}}}

\begin{figure}
  \begin{center}
    \ifthenelse{\boolean{proc}}
               {\includegraphics[width=0.25\textwidth]{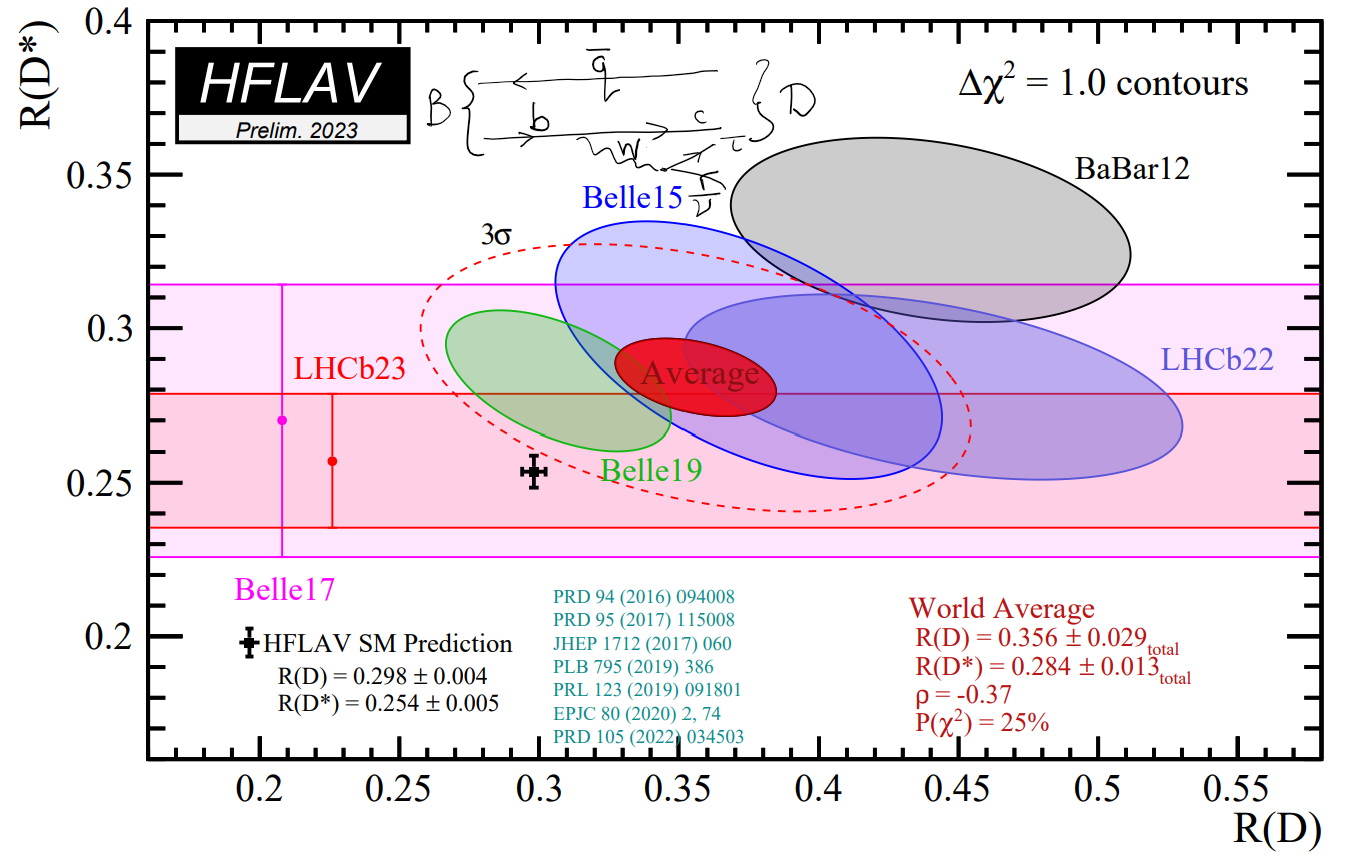}}
               {\includegraphics[width=0.5\textwidth]{anc/rd_rdstar.png}}
  \end{center}
  \caption{\label{fig:hflav}Preliminary HFLAV 2023 world average of $R(D)$ and
    $R(D\ast)$, along with the HFLAV SM 
    predictions. Figure taken from Ref.~\protect\cite{HFLAV}.
    The horizontal bands are 68$\%$ confidence level bands, whereas the
    filled ellipses are 39$\%$ bands. The dashed ellipse shows the locus of
    $p-$value where a univariate Gaussian distribution would be at $3 \sigma$
    from the maximum. The 39$\%$ confidence level (CL) world average is shown
    as the filled red 
    ellipse. Near the top of the figure, an example leading order Feynman
    diagram has been sketched in order to
    show a leading SM contribution to $B\rightarrow D \tau \bar \nu$ decay.} 
\end{figure}

Measurements of ratios of the branching ratios
\begin{equation}
  R(D{(\ast)}) = \frac{B \rightarrow D^{(\ast)} \tau \bar \nu}{B \rightarrow
    D{(\ast)} \ell \bar \nu},
\end{equation}
where $\ell \in \{ e, \mu\}$
are of interest because they provide a test of lepton flavour universality
(LFU) of
the couplings of $W^\pm$ bosons in the Standard Model (SM). A recent estimate
of the world 
average by the 
HFLAV collaboration put the joint determination of $R(D)$ and $R(D\ast)$ at a
tension with SM 
predictions at the 3.2$\sigma$ level~\cite{HFLAV}. This tension is called the
$b \rightarrow c \tau \nu$ anomaly\footnote{We note that each observable
implicitly includes an average over different charge decays.}
and has led various authors to suggest that quantum fields associated with
TeV-scale leptoquarks, charged Higgs' or
$W^\prime$ particles may be contributing to the process.
We show a preliminary 2023 HFLAV combination
in Fig.~\ref{fig:hflav}, where some measurements have been updated as compared
to 
the preceding official HFLAV combination~\cite{HFLAV2}.

\begin{figure}
  \begin{center}
    \ifthenelse{\boolean{proc}}
               {\includegraphics[width=0.25 \textwidth]{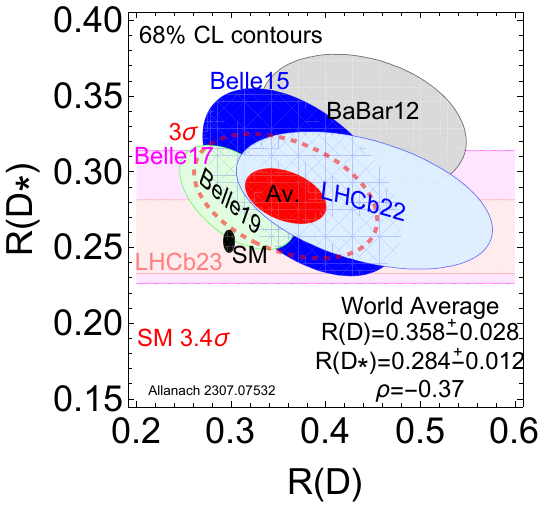}
                 \includegraphics[width=0.25 \textwidth]{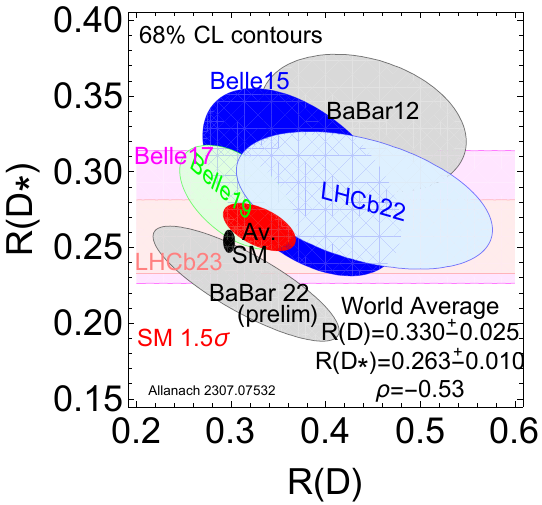}}
               {\includegraphics[width=0.45 \textwidth]{anc/rdrds_old}
                 \includegraphics[width=0.45 \textwidth]{anc/rdrds_new}}
  \end{center}
  \caption{\label{fig:myfit}Our determination of the world average of $R(D)$ and
    $R(D\ast)$ and the SM 
    prediction (left panel) excluding and (right panel) including the 2022
    semi-leptonic tag (SL) 
    preliminary measurements from BaBar data. The horizontal bands and error
    ellipses are 
    68$\%$ confidence 
    level bands, except for the dashed ellipse; this shows the locus of
    iso-$p-$value where a univariate Gaussian distribution would be at $3
    \sigma$ from the maximum.}
\end{figure}

Other measurements have recently come to our attention: a 2022
simultaneous determination of $R(D)$ and $R(D\ast)$ 
based on 
semi-leptonic decays\footnote{This is \emph{not} an official BaBar
result~\cite{lusiani}; that is currently in progress but not yet complete.} of the tagged $B$ meson in BaBar data~\cite{thesis}. The
2012 BaBar determination 
featured in Fig.~\ref{fig:hflav} was based instead on hadronic decays of the
tagged $B$ meson.
The 2022 measurements are reported as
\begin{equation}
  R(D)=0.316\pm 0.062 \pm 0.019, \qquad
  R(D\ast)=0.226 \pm 0.022 \pm 0.012, \qquad
  \rho = -0.82 \label{babar:2022}.
\end{equation}
It is our purpose here to augment the world average by these measurements.
We shall perform a more approximate job
than has HFLAV\@. In particular, we have not correlated any systematic errors; 
otherwise we have added all errors in quadrature. Correlations
between $R(D)$ and $R(D\ast)$ were taken into account, though.
We now check the level of agreement we obtain with HFLAV's more accurate
calculation. We 
present the results of 
our calculation, based on the same data, in Fig.~\ref{fig:myfit}.
We display 68$\%$ confidence level
contours consistently throughout (except for the 3$\sigma$ one).
The results are in fact
similar (although one should note the differences in appearance coming from
choosing somewhat different confidence levels and aspect ratio),
showing small differences due to the further approximations of 
the latter combination. The central values differ by a percent relative deviation
between the two determinations.
HFLAV´s determination holds that the tension between data and the SM is at an
equivalent univariate Gaussian distribution value (EUGDV) of 3.2$\sigma$,
whereas our calculation yields 3.4$\sigma$. 
We obtain a $p-$value of consistency of the measurements is {}.26 rather than
{}.25. 
We judge that these results are close enough to validate our approximations.

Our
determination of the world 
average \emph{now including the 2022 measurements of the semi-leptonic tagged
BaBar data} is
\begin{equation}
  R(D)=0.339\pm 0.025, \qquad
  R(D\ast)=0.263 \pm 0.01, \qquad
  \rho = -0.53, \label{combabar:2022}.  
\end{equation}
which we calculate to be in only 1.5$\sigma$ EUGDV tension with the HFLAV 2023
SM 
prediction. We note here that the 2022 semi-leptonic tag result is in some
tension 
with the other measurements, reducing the $p-$value of the measurements to an
EUGDV of 2.7$\sigma$.
The calculations used to produce Fig.~\ref{fig:myfit} are included in the
ancillary 
information of the {\tt arXiv} version of this manuscript.

Since the combined world average no longer calls strongly for new physics
effects in $b 
\rightarrow c \tau \nu$ decays (although we note the remaining tensions
between measurements and that there is still some room
for new physics effects), we move on to $b \rightarrow s \ell^+\ell^-$ 
decays.

\ifthenelse{\boolean{proc}}{}{\section{$b \rightarrow s \ell^+ \ell^-$ processes}}
\begin{figure}
  \begin{center}
  \includegraphics[width=0.225 \textwidth]{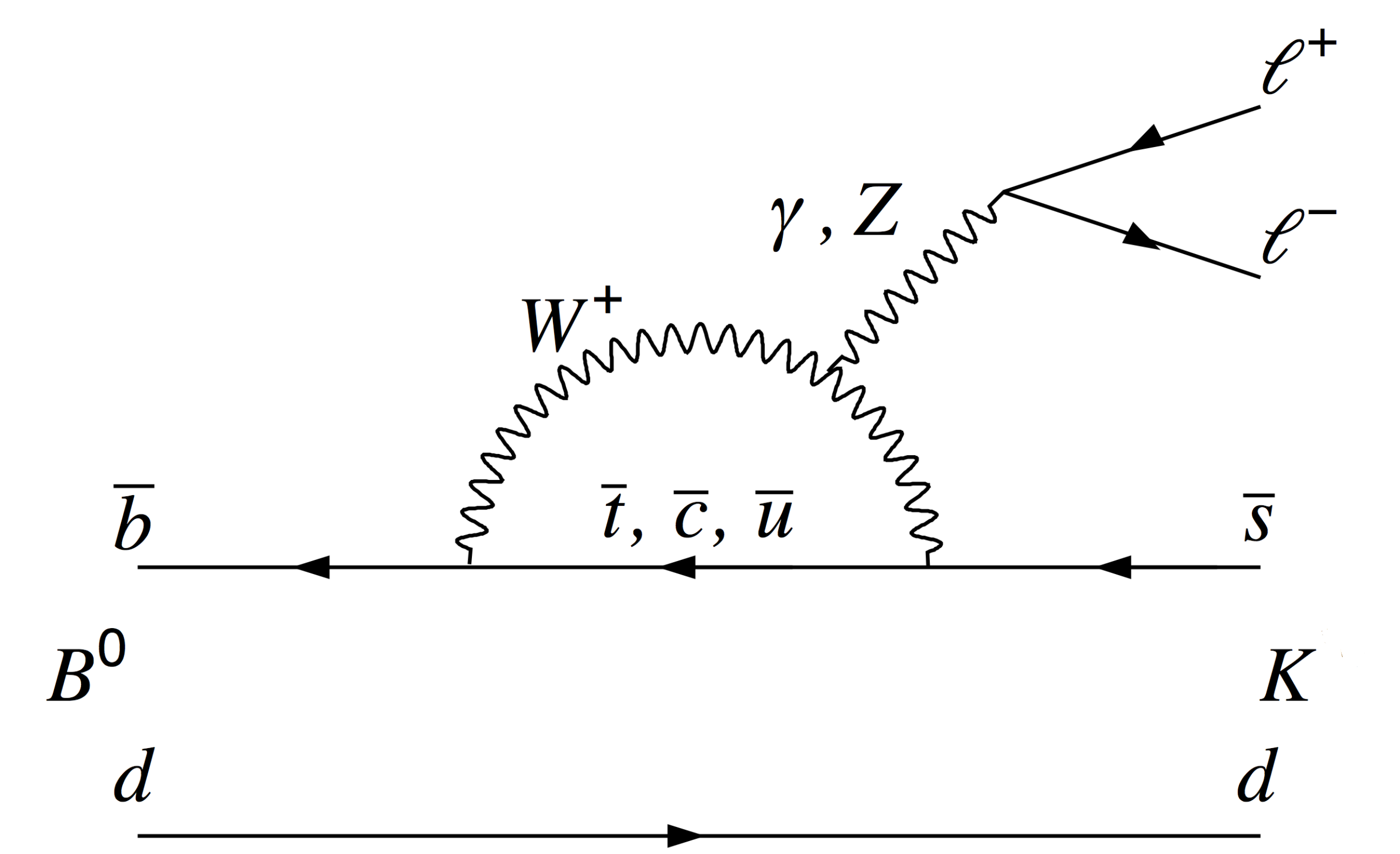}
  \end{center}
  \caption{\label{fig:peng}An example of a penguin diagram
    contributing to the   $\bar B \rightarrow \bar K^{\ast 0} \ell^+ \ell^-$
    process.} 
\end{figure}
In the SM, the dominant diagrams contributing to $B\rightarrow K^\ast \ell^+ \ell^-$
are one-loop electroweak box or penguin
contributions such as the penguin contribution shown in Fig.~\ref{fig:peng}. These decays are
also 
useful for testing LFU\@. The amplitude is suppressed by the loop, by the fact
that it is an electroweak process rather than a strong process, and by small
CKM entries. This has the result that the branching ratio for $BR(B \rightarrow
K^{(\ast)}\ell^+\ell^-)$ is of order $10^{-6}$ or less. Since the electron- and
muon- masses are much smaller than the $B$ meson mass, they can be 
approximated as massless. This approximation leads to the prediction
$BR(B\rightarrow 
K^{(\ast)} 
\mu^+\mu^-) = BR(B \rightarrow K^{(\ast)} e^+e^-)$, which holds to high
precision except at low $q^2:=m_{ll}^2$.

In practice, Feynman diagrams such as the one in Fig.~\ref{fig:peng} receive
QCD corrections. These can be parameterised by form factors (scalar
functions of $q^2$).
The predictions for the rare decay $B \rightarrow M \ell^+ \ell^-$ then are
written in terms of form factors multiplied by kinematic variables and
pre-factors. 
The form factors come
in two categories: \emph{local} and \emph{non-local} form factors. For the
local form factors, one can interpolate lattice results which are valid at
high $q^2$ (and therefore low $M$ recoil) and light cone sum rule at low
$q^2$. The non-local form factors currently have no precise lattice
estimates. Several of the SM predictions that are used to produce fits that we
shall show use QCD factorisation
plus an ad-hoc parameterisation of the long distance contribution which is fit
to data (the dominant contribution comes from charm loops). We will also show
several SM predictions from the {\tt
  EOS}~\cite{Gubernari:2022hxn,Gubernari:2023puw} approach. Here, the light 
cone operator product expansion at $q^2<0$ is interpolated/extrapolated to
various 
measurements of branching ratios and angular distributions in decays, which
are made at a $q^2$ value of the $J/\psi$ resonance mass squared,
$M_{J/\psi}^2$.
Dispersion relations bound the coefficients of a polynomial expansion in terms
of a specific kinematic variable, allowing truncations to finite order and a
resulting fit of them to a finite number of measurements.
In the {\tt EOS} approach~\cite{Gubernari:2022hxn,Gubernari:2023puw}, the dominant remaining uncertainty
on the relevant predictions typically comes from uncertainties in the local
form factors. 

\begin{figure}
  \begin{center}
    \ifthenelse{\boolean{proc}}
      {\includegraphics[width=0.25 \textwidth]{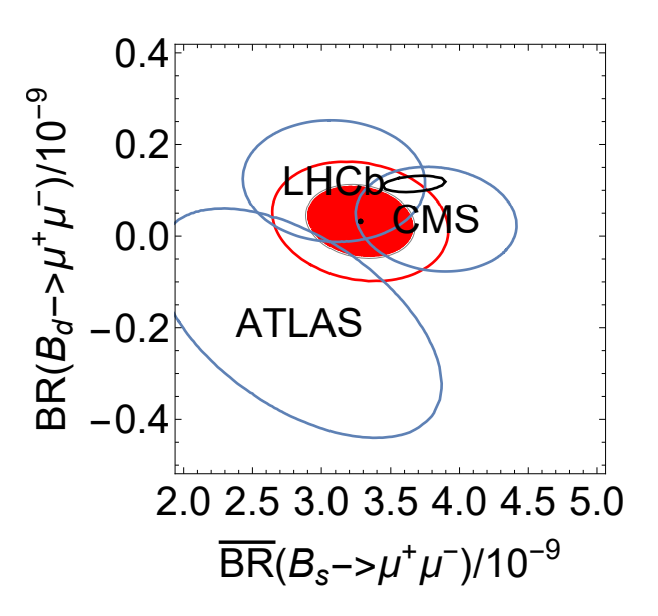}
        \includegraphics[width=0.25 \textwidth]{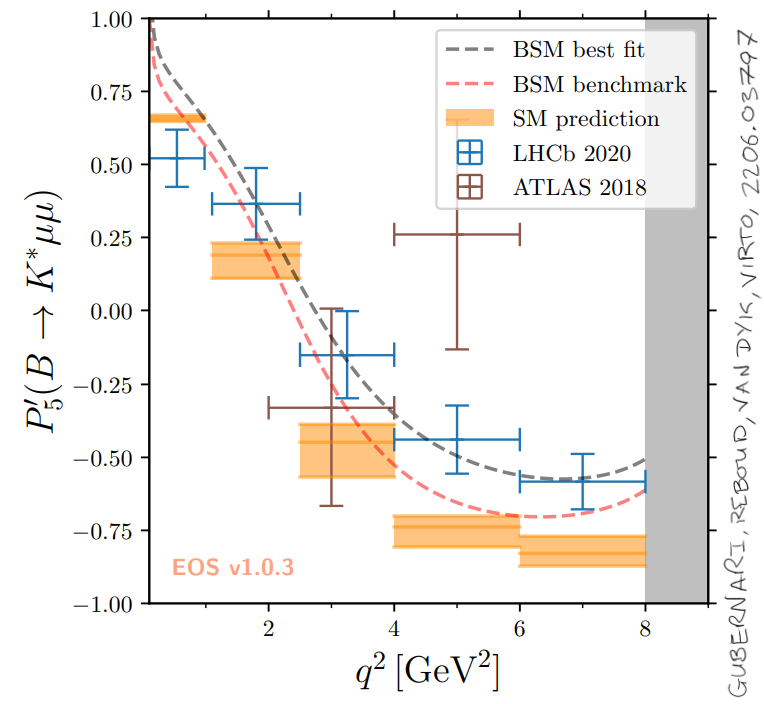}\\
        \includegraphics[width=0.25 \textwidth]{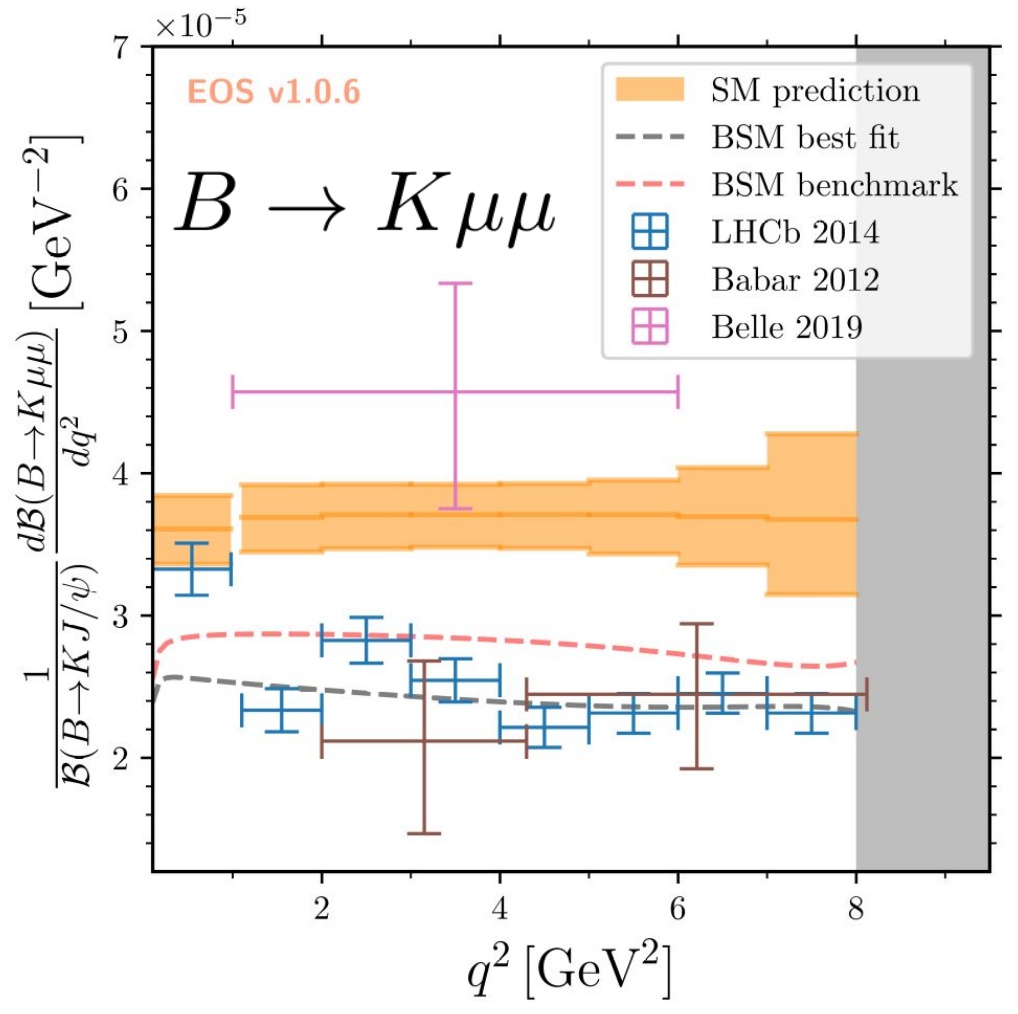}
        \includegraphics[width=0.25 \textwidth]{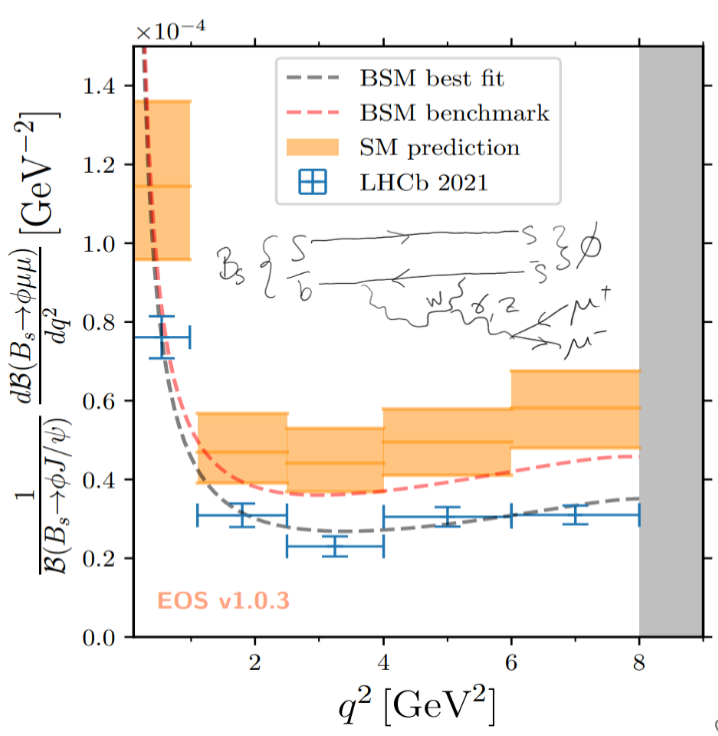}}
      {\includegraphics[width=0.45 \textwidth]{anc/bsmumu_comb.png}
        \includegraphics[width=0.45 \textwidth]{anc/p5pnew.png}\
        \includegraphics[width=0.45 \textwidth]{anc/br_rk.png}
        \includegraphics[width=0.45 \textwidth]{anc/bs_to_phimumu.png}}
  \end{center}
  \caption{\label{fig:SMpreds} Various SM predictions and measurements of
    $bs\mu\mu$ processes. (top left) $BR(B_s\rightarrow \mu^+\mu^-)$: the SM
    prediction is from Ref.~\cite{Feldmann:2022ixt} and the combination of data in red is from
    Ref.~\cite{Allanach:2022iod}. Empty ellipses show the 95$\%$ CL constraint
    in each case.
    (top right) shows a certain  angular
  distribution variable, $P_5^\prime$, in $B\rightarrow K^\ast \mu^+ \mu^-$
  decays. (bottom left) shows $BR(B_s \rightarrow \phi \mu^+\mu^-)$ as a
  function of $q^2$ and (bottom right) shows $BR(B \rightarrow K \mu^+
  \mu^-)$ decay along with a hand-drawn leading order SM contribution. In the
  latter three sub-plots, SM predictions are made by 
     {\tt EOS}~\cite{Gubernari:2022hxn,Gubernari:2023puw}.}
\end{figure}

The form factors for the SM prediction of $BR(B_s\rightarrow \mu^+\mu^-)$ are
known accurately by lattice computations.
We display this branching ratio, jointly measured with $BR(B_s\rightarrow
\mu^+\mu^-)$ in Fig.~\ref{fig:SMpreds}, along with some other $b \rightarrow s
\mu^+ \mu^-$ observables. 
Fig.~\ref{fig:SMpreds} shows that several of the observables involving the $b
\bar s \mu^+ \mu^-$ vertex display tensions between SM predictions and
measurements in several bins of $q^2$. One observes varying degrees of
tension in the SM predictions and data in each observable, although in the
$CP-$untagged $\overline{BR}(B_s \rightarrow \mu^+\mu^-)$ decay, the tension is
only mild at 1.6$\sigma$ of an EUGDV\@. Many authors have interpreted the
tensions, dubbed the 
`$b\rightarrow s \mu\mu$ anomalies', as requiring contributions from processes
involving 
 quantum fields beyond the SM. 

\ifthenelse{\boolean{proc}}{}{\subsection{Lepton flavour universality ratios}}
LHCb has published a reanalysis~\cite{LHCb:2022qnv} of the measurements of
$R_K$ and 
$R_{K^\ast}$, where
\begin{equation}
  R_{X}(q^2)=\frac{\int_{q^2_{min}}^{q^2_{max}} BR(B\rightarrow
    X\mu^+\mu^-(q^2))}{\int_{q^2_{min}}^{q^2_{max}} BR(B\rightarrow X e^+e^-(q^2))}.
\end{equation}
$q^2_{min}$ and $q^2_{max}$ are the two extreme values of $q^2$ in the
particular $q^2$
bin under consideration. `low-$q^2$' corresponds to $q^2 \in (0.1 ,\ 1,1 )$
GeV$^2$ whereas `central-$q^2$' corresponds to $q^2 \in (1.1,\ 6.0)$ GeV$^2$.
These measurements show no significant deviation from the SM. 
\begin{figure}
  \begin{center}
    \ifthenelse{\boolean{proc}}
               {\includegraphics[width=0.25 \textwidth]{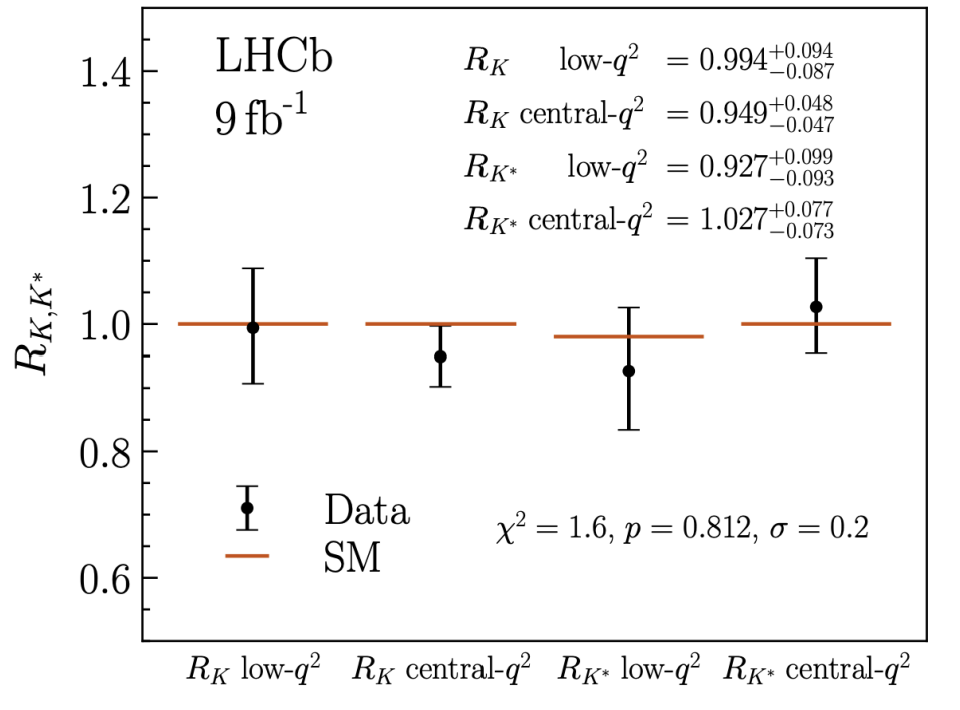}\\
                 \includegraphics[width=0.25 \textwidth]{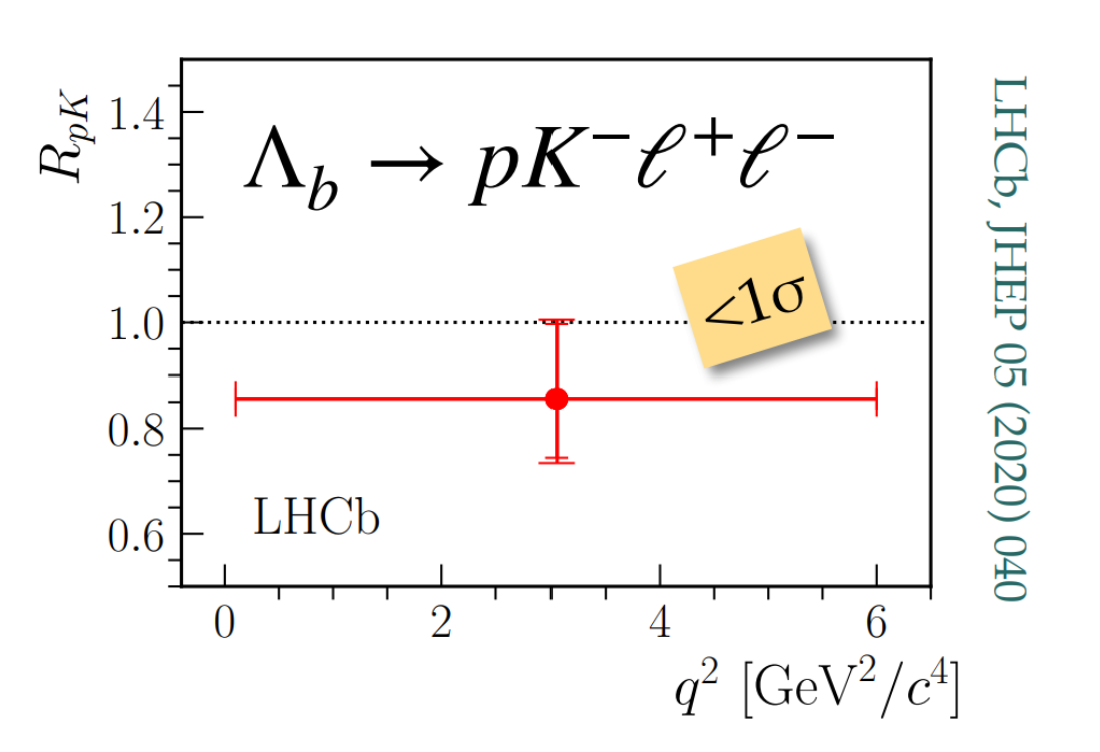}
                 \includegraphics[width=0.25 \textwidth]{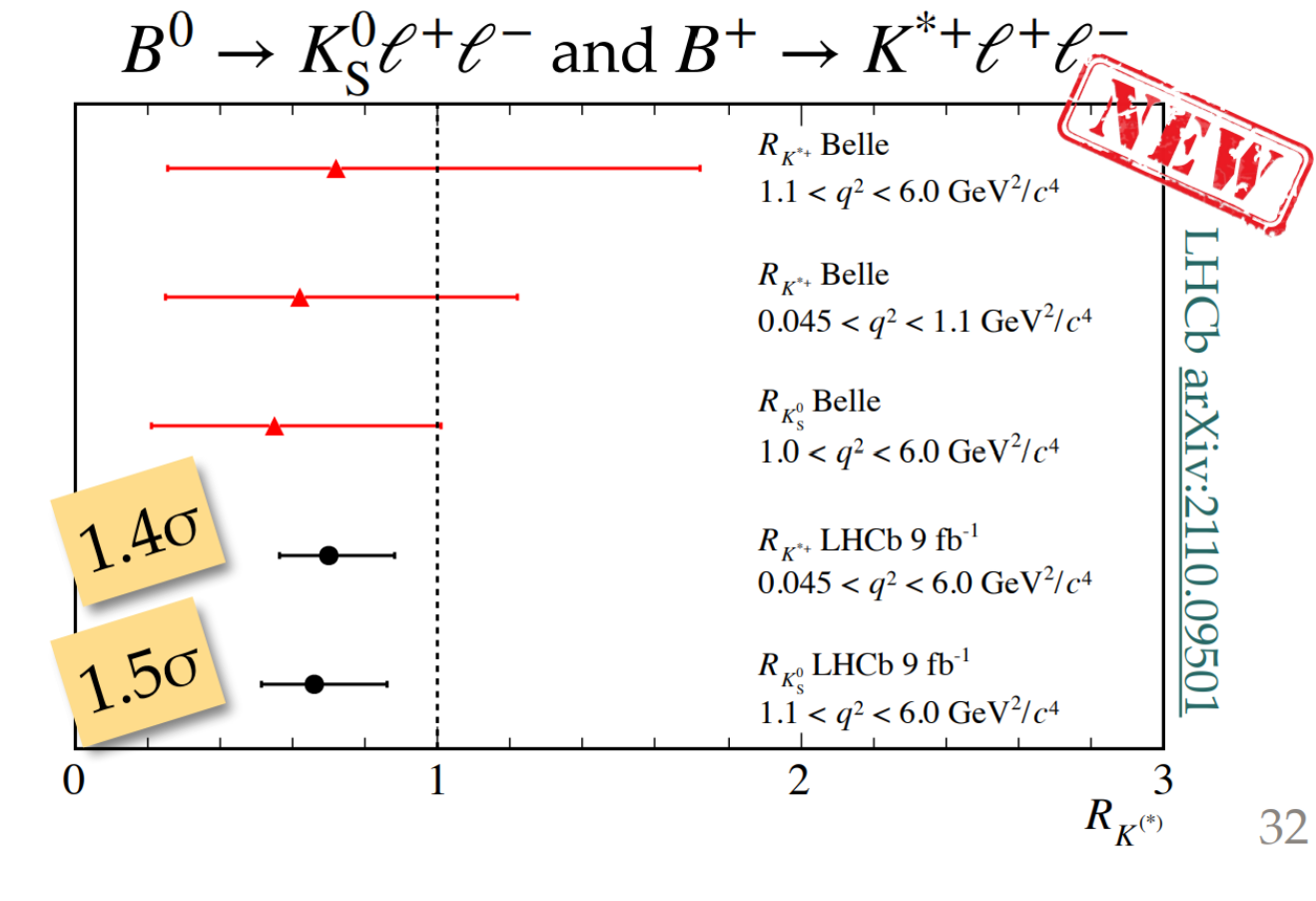}}
               {\includegraphics[width=0.45 \textwidth]{anc/new_rs}\\
                 \includegraphics[width=0.45 \textwidth]{anc/lfu1}
                 \includegraphics[width=0.45 \textwidth]{anc/lfu2}}
  \end{center}
  \caption{\label{fig:rk}(top) Recent LHCb measurements of $R_K$ and
    $R_{K^\ast}$~\cite{LHCb:2022qnv}. (bottom-left) LHCb measurement of
    $R_{pK}$. (bottom-right) Belle measurements of $R_K$ and $R_{K^\ast}$ and
    LHCb measurements of other LFU ratios.} 
\end{figure}
This is a relatively new development, and indicates that, if there is new
physics in $b \rightarrow s \mu\mu$ transitions, as indicated above, there may
also be associated new physics in $b \rightarrow s e^+e^-$ in order that
$R_K$ and $R_{K^\ast}$ may be close to the lepton flavour universality limit.

\ifthenelse{\boolean{proc}}{}{\section{Weak effective theory fits}}
We now interpret the new physics effects in terms of the Weak Effective Theory
(WET), which parameterises the effects of integrated out heavy new physics
fields in terms of new 
effectively non-renormalisable operators at the weak scale or below. 
Two such operators that have been shown to substantially ameliorate
the fit to the 
$b\rightarrow s \mu^+ \mu^-$ anomalies are
\begin{equation}
  {\mathcal L} = \ldots + N \left( C_{9\mu}^{NP} (\bar b \gamma^\alpha P_L s) (\bar \mu
  \gamma_\alpha \mu)  + C_{10\mu}^{NP}(\bar b \gamma^\alpha P_L s) (\bar \mu
  \gamma_\alpha \gamma_5 \mu)  + H.c.\right), \label{c9mu}
\end{equation}
$N:=4 G_F e^2 |V_{ts}| / (16 \pi^2\sqrt{2})$ is a normalising constant, where
$G_F$ is the Fermi decay constant, $e$ the electromagnetic gauge coupling and
$V_{ij}$ the entries of the CKM matrix. 
\begin{figure}
  \begin{center}
    \ifthenelse{\boolean{proc}}
               {\includegraphics[width=0.344 \textwidth]{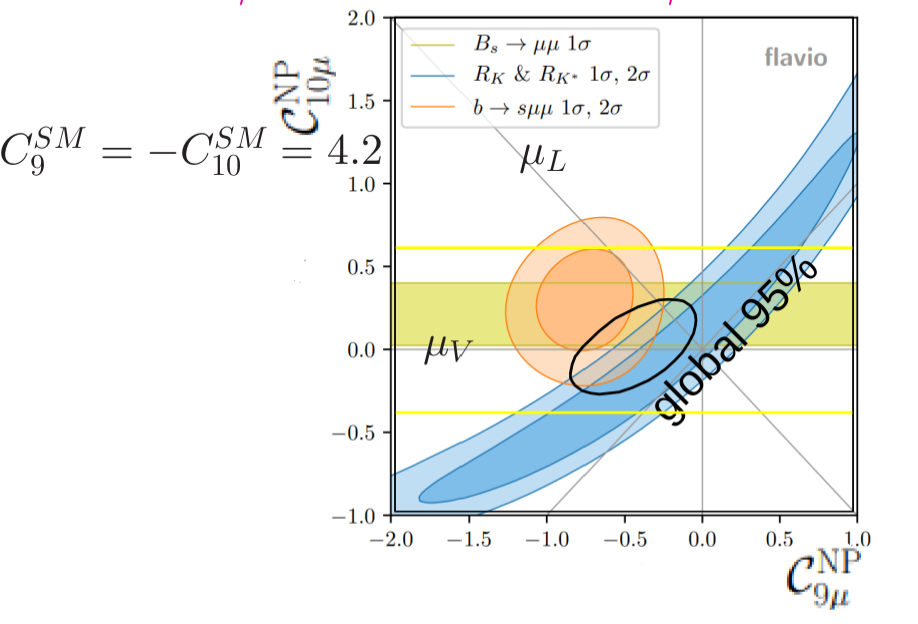}
                 \includegraphics[width=0.25 \textwidth]{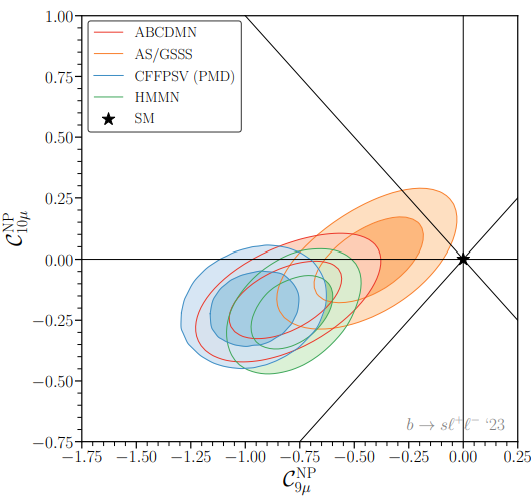}}
               {\includegraphics[width=0.55 \textwidth]{anc/c9muc10mu}
                 \includegraphics[width=0.4 \textwidth]{anc/comparison}}
  \end{center}
  \caption{\label{fig:ffits} Fits of two non-zero WET operators to $b
    \rightarrow s \ell \ell$ anomaly data. The inner (outer) coloured bands
    show the 
    68$\%$ ($95\%$ CL) regions. (left) {\tt flavio}
    fits~\cite{Greljo:2022jac}.  `$b \rightarrow s \mu\mu$'
    contains observables such as $B \rightarrow K^{(\ast)} \mu^+ \mu^-$
    branching ratios and angular 
    distributions. The global constraint using all three sets of
    observables is shown in black at the 95$\%$ CL\@. (right-panel) shows the
    global fit regions of {\tt flavio}~\cite{Greljo:2022jac} and other fitting
    groups~\cite{Alguero:2023jeh,Ciuchini:2022wbq} such as {\tt superIso}~\cite{Mahmoudi:2008tp}
    which use different sets of
    experimental data and different treatments of the SM predictions and
    theoretical uncertainties~\cite{BERNAT}.}
\end{figure}
We see the fit by the {\tt flavio} program in the left-hand panel of
Fig.~\ref{fig:ffits}.
Since the three coloured regions overlap at 95$\%$ Cl, we
conclude that there is still some compatibility with the measured values of $R_K$
and 
$R_{K^\ast}$. The figure also shows a comparison of the
fits obtained by different 
fitting groups in order to show the spread in predictions. While there is
broad agreement that the fits disagree with SM predictions\footnote{Some 
estimates in Ref.~\cite{Ciuchini:2022wbq} fit an unidentified non-perturbative
SM contribution
(that mimics a $q^2-$dependent lepton-family universal $C_9$) in tandem with the
new physics WET  operators.
As argued in Ref.~\cite{Isidori:2023unk}, a similar non-perturbative effect
cannot explain the 2$\sigma$ deficit in the
$BR(B \rightarrow X_s 
\mu^+ \mu^-)$ high $q^2-$bin, which is compatible with the low
$q^2-$deficits.}, there are some quantitative differences visible between the
fits. 

\begin{figure}
  \begin{center}
    \ifthenelse{\boolean{proc}}
               {\includegraphics[width=0.25 \textwidth]{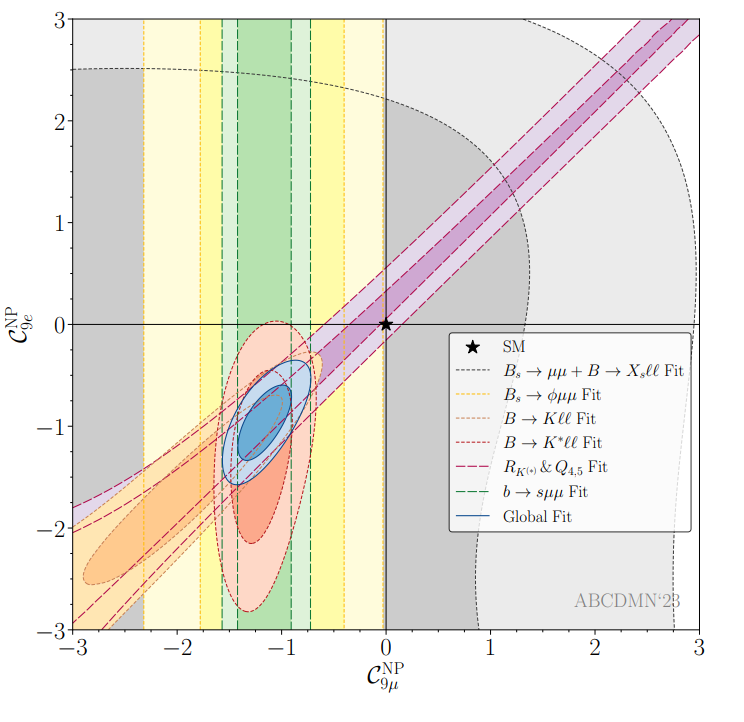}}
               {\includegraphics[width=0.4 \textwidth]{anc/c9muc9e}}
  \end{center}
  \caption{\label{fig:fit} Fits of $C_{9\mu}^{NP}$ and $C_{9e}^{NP}$ to $b
    \rightarrow s \ell \ell$ anomaly data. The inner (outer) coloured bands show the
    68$\%$ ($95\%$ CL) regions, respectively. Figure taken from
    Ref.~\cite{Alguero:2023jeh}.} 
\end{figure}

Now, we consider also turning on a new physics operator involving di-electron
pairs
\begin{equation}
  {\mathcal L} = \ldots + N  C_{9e}^{NP} (\bar b \gamma^\alpha P_L s) (\bar e
  \gamma_\alpha e) + H.c. \label{c9e}
\end{equation}
We display a joint WET fit to non-zero $C_{9\mu}^{NP}$ and $C_{9e}^{NP}$ in
Fig.~\ref{fig:fit}, where we see the preference for $C_{9e}^{NP} \neq 0$. Overall,
the fit has a pull away from the SM limit (the origin), equivalent to
a UEDGV of 5.2$\sigma$, a significant
change for only two fitted parameters.

We now turn to some simple bottom-up models which can generate the non-zero
WET operators which we mention above and which were found to significantly
ameliorate SM predictions. 

\begin{figure}
  \begin{center}
    \includegraphics[width=0.45 \textwidth]{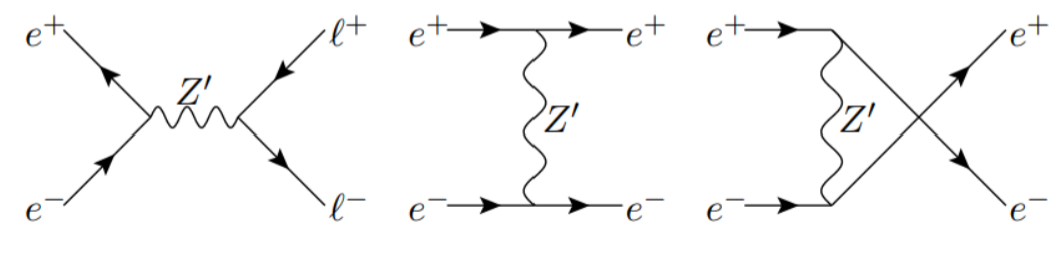}
  \end{center}
  \caption{\label{fig:lep} Leading Feynman diagrams of $Z^\prime$
    contributions to di-lepton production at the LEP collider.} 
\end{figure}

\begin{figure}
  \begin{center}
    \ifthenelse{\boolean{proc}}    
               {\includegraphics[width=0.25 \textwidth]{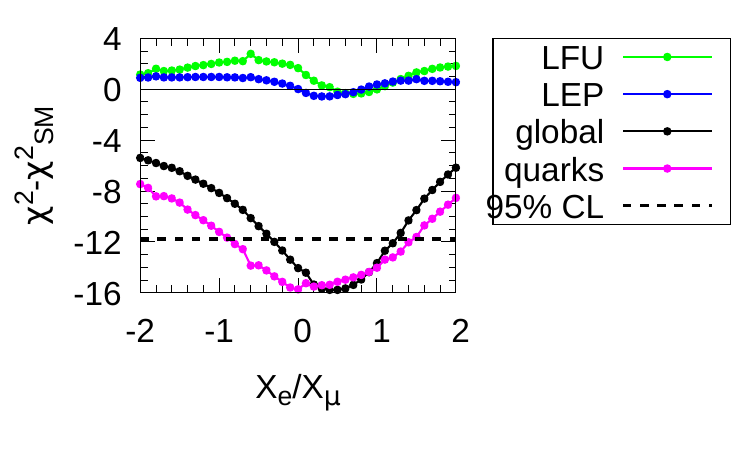}}
               {\includegraphics[width=0.5 \textwidth]{anc/chi_squared}}
  \end{center}
  \caption{\label{fig:chisq} Two parameter fits of a set of $Z^\prime$ models
    to $b 
    \rightarrow s \ell \ell$ anomaly data. Figure taken from
    Ref.~\cite{Allanach:2023uxz}. Regions where the `global' curve is below the dashed line are compatible
    with data globally, to 95$\%$ CL, assuming the hypothesis of the line of
    models.} 
  \end{figure}

\ifthenelse{\boolean{proc}}{}{\section{Simple models}}
We examine simple $Z^\prime$ models with a spontaneously broken $U(1)_X$ gauge
extension of the 
SM~\cite{Allanach:2023uxz}, 
where
\begin{equation}
  3B_3 - \left(X_e L_e + X_\mu L_\mu + [3 - X_e - X_\mu]L_\tau\right),
  \label{b3mL}
\end{equation}
i.e.\ where $B_3$ is third family baryon number, $L_e$ is electron number,
$L_\mu$ is muon number and $X_e$ and $X_\tau$ are arbitrary integer
parameters. Such an assignment is anomaly free if three 
right-handed neutrinos augment the SM chiral fermion content.
Once the $Z^\prime$ is coupled to di-electron pairs as would be implied by
$X_e \neq 0$, one should apply bounds 
from LEP which come from differential measurements of scattering to di-lepton
pairs. These do not show significant tensions with SM
predictions~\cite{Falkowski:2015krw} and so they bound the contributions coming
from Feynman diagrams such as those shown in Fig.~\ref{fig:lep}.
(\ref{b3mL})
allows one to interpolate between $Z^\prime$ models
which couple to charged 
leptons via muons only, to those which couple equally to electrons and muons.
We display a set of global fits to different values of $X_e$, for $X_\mu=10$,
in Fig.~\ref{fig:chisq}.
  The figure shows that a significant improvement on $\chi^2$ is obtained as
  compared to the SM for $X_e/X_\mu \approx 1/2$, although any value in the
  range $(-0.4,1.3)$ is within the 95$\%$ CL\@. We also see that the limits of
  zero coupling of the $Z^\prime$ boson to di-electron pairs and equal
  coupling to di-electron and di-muon pairs are fit more-or-less equally well,
  with only an insignificant 
  difference in best-fit $\chi^2$ value of 0.7 between them. 

\begin{figure}
    \begin{center}
      \ifthenelse{\boolean{proc}}
                 {\includegraphics[width=0.15 \textwidth]{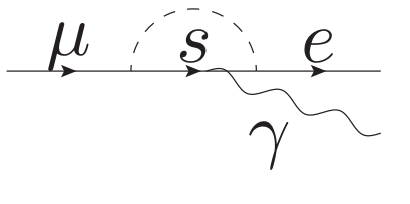}}
                 {\includegraphics[width=0.25 \textwidth]{anc/meg}}
  \end{center}
  \caption{\label{fig:meg} A scalar LQ's (that has been fit
    to $b \rightarrow s \ell^+ \ell^-$ anomalies and therefore has a coupling
    to the strange quark) contribution to $\mu \rightarrow e
    \gamma$. There is another contribution with $s \rightarrow b$.}
\end{figure}

  We anticipate that requiring a single leptoquark (LQ) to couple to electrons
  and muons with a similar 
  strength in order to fit $R_K$ and $R_{K^\ast}$ will generically lead to
  contravention of the 
  $BR(\mu \rightarrow e \gamma)<4.2 \times 10^{-13}$ (90$\%$ CL) bound from
  the MEG collaboration~\cite{MEG:2016leq} through processes such as the one
  shown in Fig.~\ref{fig:meg}\footnote{$\mu \rightarrow e$ conversion also
  provides very strong constraints upon such a LQ.}.
However, Ref.~\cite{Allanach:2022iod} found that LQs that do \emph{not} couple to di-electron pairs, can
provide an improvement on SM predictions similar to that of
the `$B_3-L_2$' model with $X_\mu=3, X_e=0$.

\ifthenelse{\boolean{proc}}{}{\section*{Acknowledgements}
This work has been partially supported by STFC HEP Theory Consolidated grant
ST/000694. We thank other members of the Cambridge Pheno Working group for
helpful discussions and Lepton-Photon 2023 for facilitating the talk.}

\bibliographystyle{JHEP-2}       
\bibliography{article}   

\providecommand{\href}[2]{#2}\begingroup\raggedright\begin{thebibliography}{10}

\bibitem{HFLAV}
{\bf Heavy Flavor Averaging Group, HFLAV} Collaboration, Y.~S. Amhis {\em
  et.~al.}, {\it Preliminary average of $r(d)$ and $r(d\ast)$ for winter 2023},
  .
  https://hflav-eos.web.cern.ch/hflav-eos/semi/winter23\_prel/html/RDsDsstar/RDRDs.html.

\bibitem{HFLAV2}
{\bf Heavy Flavor Averaging Group, HFLAV} Collaboration, Y.~S. Amhis {\em
  et.~al.}, {\it {Averages of b-hadron, c-hadron, and \ensuremath{\tau}-lepton
  properties as of 2021}},  {\em Phys. Rev. D} {\bf 107} (2023), no.~5 052008
  [\href{http://arXiv.org/abs/2206.07501}{{\tt 2206.07501}}].

\bibitem{lusiani}
A.~Lusiani, {\it private communication},  July, 2023.

\bibitem{thesis}
Y.~Li, {\em {Search for Beyond Standard Model Physics at Babar}}.
\newblock PhD thesis, Caltech, 2022.
\newblock https://resolver.caltech.edu/CaltechTHESIS:05232022-144829107.

\bibitem{Gubernari:2022hxn}
N.~Gubernari, M.~Reboud, D.~van Dyk and J.~Virto, {\it {Improved theory
  predictions and global analysis of exclusive $b \to s\mu^+\mu^-$ processes}},
   {\em JHEP} {\bf 09} (2022) 133 [\href{http://arXiv.org/abs/2206.03797}{{\tt
  2206.03797}}].

\bibitem{Gubernari:2023puw}
N.~Gubernari, M.~Reboud, D.~van Dyk and J.~Virto, {\it {Dispersive Analysis of
  $B\to K^{(*)}$ and $B_s\to \phi$ Form Factors}},
  \href{http://arXiv.org/abs/2305.06301}{{\tt 2305.06301}}.

\bibitem{Feldmann:2022ixt}
T.~Feldmann, N.~Gubernari, T.~Huber and N.~Seitz, {\it {Contribution of the
  electromagnetic dipole operator O7 to the
  B\textasciimacron{}s\textrightarrow{}\ensuremath{\mu}+\ensuremath{\mu}- decay
  amplitude}},  {\em Phys. Rev. D} {\bf 107} (2023), no.~1 013007
  [\href{http://arXiv.org/abs/2211.04209}{{\tt 2211.04209}}].

\bibitem{Allanach:2022iod}
B.~Allanach and J.~Davighi, {\it {The Rumble in the Meson: a leptoquark versus
  a Z' to fit b \textrightarrow{} s\ensuremath{\mu}$^{+}$\ensuremath{\mu}$^{-}$
  anomalies including 2022 LHCb $ {R}_{K^{\left(\ast \right)}} $
  measurements}},  {\em JHEP} {\bf 04} (2023) 033
  [\href{http://arXiv.org/abs/2211.11766}{{\tt 2211.11766}}].

\bibitem{LHCb:2022qnv}
{\bf LHCb} Collaboration, R.~Aaij {\em et.~al.}, {\it {Test of lepton
  universality in $b \rightarrow s \ell^+ \ell^-$ decays}},  {\em Phys. Rev.
  Lett.} {\bf 131} (2023), no.~5 051803
  [\href{http://arXiv.org/abs/2212.09152}{{\tt 2212.09152}}].

\bibitem{Greljo:2022jac}
A.~Greljo, J.~Salko, A.~Smolkovi\v{c} and P.~Stangl, {\it {Rare b decays meet
  high-mass Drell-Yan}},  {\em JHEP} {\bf 05} (2023) 087
  [\href{http://arXiv.org/abs/2212.10497}{{\tt 2212.10497}}].

\bibitem{Alguero:2023jeh}
M.~Alguer\'o, A.~Biswas, B.~Capdevila, S.~Descotes-Genon, J.~Matias and
  M.~Novoa-Brunet, {\it {To (b)e or not to (b)e: No electrons at LHCb}},
  \href{http://arXiv.org/abs/2304.07330}{{\tt 2304.07330}}.

\bibitem{Ciuchini:2022wbq}
M.~Ciuchini, M.~Fedele, E.~Franco, A.~Paul, L.~Silvestrini and M.~Valli, {\it
  {Constraints on lepton universality violation from rare B decays}},  {\em
  Phys. Rev. D} {\bf 107} (2023), no.~5 055036
  [\href{http://arXiv.org/abs/2212.10516}{{\tt 2212.10516}}].

\bibitem{Mahmoudi:2008tp}
F.~Mahmoudi, {\it {SuperIso v2.3: A Program for calculating flavor physics
  observables in Supersymmetry}},  {\em Comput. Phys. Commun.} {\bf 180} (2009)
  1579--1613 [\href{http://arXiv.org/abs/0808.3144}{{\tt 0808.3144}}].

\bibitem{BERNAT}
B.~Capdevila, {\it talk at beyond the flavour anomalies workshop},  2023.

\bibitem{Isidori:2023unk}
G.~Isidori, Z.~Polonsky and A.~Tinari, {\it {Semi-inclusive $b\to
  s\bar{\ell}\ell$ transitions at high $q^2$}},
  \href{http://arXiv.org/abs/2305.03076}{{\tt 2305.03076}}.

\bibitem{Allanach:2023uxz}
B.~Allanach and A.~Mullin, {\it {Plan B: New ${Z^\prime}$ models for
  $b\rightarrow sl^+l^-$ anomalies}},
  \href{http://arXiv.org/abs/2306.08669}{{\tt 2306.08669}}.

\bibitem{Falkowski:2015krw}
A.~Falkowski and K.~Mimouni, {\it {Model independent constraints on four-lepton
  operators}},  {\em JHEP} {\bf 02} (2016) 086
  [\href{http://arXiv.org/abs/1511.07434}{{\tt 1511.07434}}].

\bibitem{MEG:2016leq}
{\bf MEG} Collaboration, A.~M. Baldini {\em et.~al.}, {\it {Search for the
  lepton flavour violating decay $\mu ^+ \rightarrow \mathrm {e}^+ \gamma $
  with the full dataset of the MEG experiment}},  {\em Eur. Phys. J. C} {\bf
  76} (2016), no.~8 434 [\href{http://arXiv.org/abs/1605.05081}{{\tt
  1605.05081}}].

\end{thebibliography}\endgroup

\end{document}